\documentclass[journal]{IEEEtran}
\ifCLASSINFOpdf
\else
\fi
\usepackage{bbm}
\usepackage{array}
\newcolumntype{P}[1]{>{\centering\arraybackslash}p{#1}}
\usepackage{algorithm}
\usepackage{algorithmicx}
\usepackage{algpseudocode}
\usepackage{tabu}
\usepackage{amssymb}
\usepackage{mathrsfs}
\usepackage{amsmath,cite}
\usepackage{chemarrow}
\usepackage{hyperref}
\usepackage{url}
\usepackage{color}
\usepackage{graphicx}          % Include this line if your
\usepackage{cite} 
\usepackage{multirow}
\usepackage{epstopdf}

\newtheorem{prop}{Proposition}

\newtheorem{assump}{Assumption}

\begin{document}
\title{A Full Bayesian Approach to Sparse Network Inference using Heterogeneous Datasets}

\author{Junyang Jin, Ye Yuan, and Jorge Gon\c{c}alves% <-this % stops a space
\thanks{Junyang~Jin is with Circadian Signal Transduction Group, Department of Plant Sciences, University of Cambridge. Ye~Yuan is with School of Automation, Huazhong University of Science and Technology. Jorge~Gon\c{c}alves is with the Department of Engineering, University of Cambridge and the Luxembourg Centre for Systems Biomedicine. (Corresponding author: Ye Yuan).}}% <-this % stops a space
\maketitle

\begin{abstract}
Network inference has been attracting increasing attention in several fields, notably systems biology, control engineering and biomedicine. To develop a therapy, it is essential to understand the connectivity of biochemical units and the internal working mechanisms of the target network. A network is mainly characterized by its topology and internal dynamics. In particular, sparse topology and stable system dynamics are fundamental properties of many real-world networks. In recent years, kernel-based methods have been popular in the system identification community. By incorporating empirical Bayes, this framework, which we call KEB, is able to promote system stability and impose sparse network topology. Nevertheless, KEB may not be ideal for topology detection due to local optima and numerical errors. Here, therefore, we propose an alternative, data-driven, method that is designed to greatly improve inference accuracy, compared with KEB. The proposed method uses dynamical structure functions to describe networks so that the information of unmeasurable nodes is encoded in the model. A powerful numerical sampling method, namely reversible jump Markov chain Monte Carlo (RJMCMC), is applied to explore full Bayesian models effectively. Monte Carlo simulations indicate that our approach produces more accurate networks compared with KEB methods. Furthermore, simulations of a synthetic biological network demonstrate that the performance of the proposed method is superior to that of the state-of-the-art method, namely iCheMA. The implication is that the proposed method can be used in a wide range of applications, such as controller design, machinery fault diagnosis and therapy development.
\end{abstract}


% \fi
\IEEEpeerreviewmaketitle

\section{Introduction}
 This paper is concerned with network inference problem in the case of both industrial and biological systems. In industry, communication systems are typically designed with a sparse and stable structure to reduce energy consumption and to ensure long-term operation. In biology, most networks are inherently stable, with biochemical species maintained on a normal level. Their internal connectivity is also sparse, enabling efficient working mechanisms. It follows that, for network inference, sparsity and stability are essential preliminary conditions. 

In recent years, kernel-based non-parametric system identification methods have been prevalent. Such methods have several advantages in the real-world applications. When identifying linear models, the estimation of model complexity is avoided. More importantly, by using proper kernel functions, kernel-based methods enforce system stability effectively. The kernel machine is associated with Gaussian processes~\cite{gauss,spline}. Under the Bayesian paradigm, empirical Bayes has been widely applied for the estimation of hyperparameters of kernel functions\cite{pattern,mur}. As hyperparameters control the property of system dynamics, this combined framework (KEB) greatly improves the estimation accuracy of input-output maps of target systems~\cite{nonp4,nonp5}. Kernel-based methods have been discussed in a wide range of contexts including linear continuous/discrete time systems\cite{nonp1,nonp2,nonp8}, and nonlinear NARX, NARMAX and NFI models~\cite{nonp7}. Moreover, KEB is endowed with the mechanism of Automatic Relevance Determination (ARD) and so is able to promote sparse solutions. For example, KEB has been used to solve multi-kernel selection problems\cite{nonp3}.

In network inference, a further crucial task topology detection (selection of model structure). Under the framework of KEB, the sparsity profile of ARD parameters determines network topology~\cite{nonp}, which requires accurate estimation of hyperparameters. KEB has been shown to be robust for local optimal solutions: identified models can represent the system dynamics of ground truths quite well, even if only suboptimal solutions are achieved\cite{nonp6}. 

Hyperparameter estimation becomes more challenging, however, under the context of network inference. Where the target network is very sparse, a small estimation error of the zero structure of hyperparameters can seriously degrade the reliability of inference. This can be caused by either local optima or numerical errors during implementation. Model section strategies (e.g. backward selection~\cite{nonp}) may be applied as a remedy. In real world applications, however, they have limitations: the confidence of inference cannot be evaluated and computational cost is greatly increased especially for large-scale networks.

Monte Carlo techniques (MC) provide alternative approximation inference~\cite{pattern}. They belong to stochastic approximations based on numerical sampling rather than on approximating Bayesian models analytically like KEB. Reversible jump Markov chain Monte Carlo (RJMCMC) is one of the MC approaches that was originally developed for Bayesian model selection~\cite{RJ1}. RJMCMC is able to draw samples from a distribution whose random variables are of varying dimension. RJMCMC has been applied in many research fields including optimization~\cite{RJ2,RJ3}, machine learning~\cite{RJ4,RJ5}, signal processing~\cite{RJ8}, and system identification~\cite{RJ6,RJ7}. For a network, the dimension of model parameters depends on network topology: only true links are endowed with parameters. Therefore, RJMCMC represents a promising method for network inference.

This paper combines kernel-based methods and RJMCMC to infer sparse networks. Dynamical structure functions are used to describe networks so that the information of hidden nodes can be encoded via transfer functions. As a non-parametric method, the kernel machine is applied to impose stable impulse responses of networks. RJMCMC is adopted to explore the resulting Bayesian model whose sample space consists of multiple subspaces of different dimensionality. By traversing these subspaces, RJMCMC provides a highly efficient way to infer system dynamics and to detect network topology. In particular, the effect of ADR is maximally activated by the merit of RJMCMC, thus encouraging sparse topologies. Monte Carlo simulations indicate that our method further improves inference accuracy compared with KEB. The performance improvement is greatest when inferring a synthetic biological network. Thus the contribution of our study lies in the provision of a method that is more reliable than KEB for real-world applications.


The paper is organized as follows. Section~\ref{sec:MCMC} overviews MH-within-PCG samplers and RJMCMC. Section \ref{sec:Model} introduces dynamical structure function and formulates the full Bayesian model. Section~\ref{sec:INF} discusses network inference using RJMCMC. Section~\ref{sec:Simulation} compares the method with other approaches via Monte Carlo simulations. Finally, Section~\ref{sec:Conclusion} concludes and discusses further development in this field.

\emph{Notation}: The notation in this paper is standard. $I_m$ denotes the $m\times m$ identity matrix. For $L\in{R}^{n\times n}$, $diag\{L\}$ denotes a vector which consists of diagonal elements of matrix $L$. $[L]_{ij}$ presents the $ij$th entry and $L(:,i:j)$ the  columns from $i$ to $j$. $blkdiag\{L_1,...,L_n\}$ is a block diagonal matrix. For $l\in{R}^{n}$, $diag\{l\}$ denotes a diagonal matrix whose diagonal elements come from vector $l$. $[l]_{ij}$ denotes the $j$th element of the $i$th group of $l$. $l\geq0$ means each element of the vector is non-negative.  $y(t_1:t_2)$ denotes a row vector $\left[\begin{array}{cccc}y(t_1)&y(t_1 +1)&\cdots&y(t_2)\end{array}\right]$. 

\section{Overview of Markov Chain Monte Carlo}\label{sec:MCMC}
\subsection{MH-within-PCG Sampler}
 Markov Chain Monte Carlo (MCMC) is widely applied to draw samples from probabilistic models. The samples consist of a Markov chain that asymptotically distributes as the target distribution. The samples can be used to evaluate marginal distributions and to estimate expectation of random variables, which cannot be calculated in a closed form.

Gibbs sampling and Metropolis-Hastings method (MH) are two typical MCMC techniques~\cite{mur}. They have their own strength and weakness. Gibbs samplers have a simple structure but require analytical conditional distributions. MH samplers can sample from a distribution known up to a constant but their design is more involved. In practice, Gibbs and MH samplers are often modified and combined accordingly to handle complex distributions. For example, blocked Gibbs samplers are a minor variant of traditional Gibbs samplers, which group two or more variables and sample from their joint conditional distributions so that a better convergence property is achieved ~\cite{RJ9}. Another example is the single-site updating MH sampling where only one component of the Markov state is updated at a time so that the proposal distributions can be simplified~\cite{RJ10}.

Gibbs and MH sampling can also be combined to build a more powerful and efficient sampler. If the conditional distributions of some sampling steps of a Gibbs sampler have no closed form, one can replace these steps with MH sampling schemes, leading to a hybrid sampler (MH-within-Gibbs sampler)~\cite{RJ11}. By marginalizing out certain random variables, the convergence property of a sampler can be further improved ~\cite{RJ12}. Modified Gibbs and MH-within-Gibbs samplers based on this principle are called partially collapsed Gibbs sampler (PCG) and Metropolis-Hastings within partially collapsed Gibbs sampler (MH-within-PCG), respectively~\cite{RJ11}.

Deducing a PCG from a Gibbs sampler is nontrivial because the full conditional distributions of the  sampling steps cannot be marginalized directly. Otherwise, the invariant distribution of the Markov chain may be changed. This issue was not sufficiently aware of in much of the previous research. It has been shown that some rules must be followed to preserve the invariant distribution~\cite{RJ11}. To reduce the number of conditioned random variables, the steps called marginalization, permutation and trimming are executed in sequence~\cite{RJ12}.

 Marginalization means to move components of Markov states from being conditioned on to being sampled. For example, one can replace sampling from $p(x|y,z)$ with sampling from $p(x,y|z)$ safely. Permutation means to switch the order of sampling steps. Finally, trimming is used to discard a subset of components that are not conditioned on in the next step. For instance, if the sampling step of $p(x,y|z)$ is followed by that of $p(y|x,z)$, $p(x|z)$ can be sampled instead. Nevertheless, $p(y|z)$ is not a valid replacement since $x$ is conditioned on in the next step. 

It is important to realize that sampling steps of a PCG sampler cannot be replaced by their MH counterparts directly. A MH-within-PCG sampler must be derived from the original MH-with-Gibbs sampler following the similar rules of  PCG~\cite{RJ11}. The key point is that a full MH step of a MH-with-Gibbs sampler can be replaced by a reduced MH step only if a direct draw from the conditional distribution of the reduced quantities follows up immediately~\cite{RJ11}. For
instance, the MH step to sample from $p(x|y,z)$ in a MH-with-Gibbs sampler can be replaced by the reduced MH step to sample from $p(x|z)$, followed immediately by sampling from $p(y|x,z)$. The sampling step of $p(y|x,z)$ may or may not be trimmed, depending on the next step .

\subsection{Reversible Jump Markov Chain Monte Carlo}
Traditional MCMC is used to draw samples from a distribution of random variables whose dimension is fixed. There are cases where the dimension of random variables varies. To explore such a distribution, MCMC samplers must be able to jump between parameter subspaces of different dimensionality. Reversible Jump Markov Chain Monte Carlo (RJMCMC) was designed for this purpose ~\cite{phd1,tuto}.

For a countable collection of Bayesian models $\{\mathcal{M}_k,k\in\mathbb{Z}^+\}$, each model is characterized by a parameter vector $\theta_k\in\mathbb{R}^{d_k}$, where the dimension $d_k$ may differ from model to model. The random variable to be sampled is $x=(k,\theta_k)$ which lies in the subspace $S_k=\{k\}\times \mathbb{R}^{d_k}$ given $k$. Hence, the entire parameter space is $S=\bigcup_{k\in\mathbb{Z^+}}S_k$.

Suppose $p(x)$ is the probability density function of interest. To draw samples from $p(x)$, a reversible Markov chain $\{X^t, t\in\mathbb{Z}^+\}$ is produced regarding $p(x)$ as the invariant distribution. Each Markov state $X^t$ consists of two components, $k^t$ and $\theta^t_k$ where $k^t$ is the model index and $\theta^t_k$ is the corresponding unknown model parameter. To traverse across the parameter space $S$, different types of moves are proposed, among which
only one move is executed per iteration. Theses moves are selected randomly.  Proposal distributions are carefully designed so that 'detailed balance' is achieved for each move type. The resulting transition
distribution of the Markov chain is the mixing of that of all moves. Consequently, the invariant distribution is preserved.

Let $(k,\theta)$ be the current state $X^t$ of the Markov chain where $\theta\in\mathbb{R}^{d_k}$. Based on the proposed moves, the probability to jump from the current model $k$ to the next one $k'$ is $p_{kk'}$, where $\sum_{k'}p_{kk'}=1$. If $k'=k$, only model parameters are updated in the next state. In addition, it is possible that not all the models can be reached in the next state from the current state, depending on the moves available. Given the
proposed $k'$ with probability $p_{kk'}$, $\theta'\in\mathbb{R}^{d_{k'}}$ is generated as the proposal for the model parameter. One way to generate $\theta'$ is to first produce a random quantity $U$ with the probability density
$q_{kk'}(u|\theta)$ and then map $\theta$ and $U$ to $\mathbb{R}^{d_{k'}}$. As a result, $\theta'=g_{1kk'}(\theta,U)$ where $U\in\mathbb{R}^{d_{kk'}}$ and $g_{1kk'}:\mathbb{R}^{d_k+d_{kk'}}\rightarrow\mathbb{R}^{d_{k'}}$is a deterministic map~\cite{tuto}. The proposal $X^{prop}=(k',\theta')$ is then accepted with probability $A_{kk'}(\theta'|\theta)$. If accepted, $X^{t+1}=X^{prop}$. If not, $X_{t+1}=X_t$.

For the move from $(k,\theta)$ to $(k',\theta')$ and the reverse move from $(k',\theta')$ to $(k,\theta)$, their corresponding proposals, $(\theta,U)$ and $(\theta',U')$ must have equal dimension. This is called ‘dimension matching’: $d_k+d_{kk'}=d_{k'}+d_{k'k}$~\cite{RJ1}. In addition, there must exist a deterministic map $g_{2kk'}: \mathbb{R}^{d_k+d_{kk'}}\rightarrow\mathbb{R}^{d_{kk'}}$ such that $(\theta',U')=g_{kk'}(\theta,U)=(g_{1kk'}(\theta,U),g_{2kk'}(\theta,U))$ where map $g_{kk'}$ is bijective and differentiable~\cite{tuto}.

Finally, to achieve 'detailed balance', the following equation must be satisfied~\cite{tuto}:
\begin{equation}
\begin{aligned}
&\pi(k,\theta)p_{kk'}q_{kk'}(U|\theta)A_{kk'}\\
&=\pi(k',\theta')p_{k'k}q_{k'k}(U'|\theta')A_{k'k}\left|\frac{\partial g_{kk'}(\theta,U)}{\partial \theta\partial U}\right|.\\
\end{aligned}
\end{equation}
where
\begin{equation}
\begin{aligned}
\theta' = g_{1kk'}(\theta,U)~\text{and}~
U' = g_{2kk'}(\theta,U).
\end{aligned}
\end{equation}
As a result, the acceptance probability equates to:
\begin{equation}
\begin{aligned}
&A_{kk'}(\theta'|\theta)\\
&=\min\left\{1,\frac{\pi(k',\theta')p_{k'k}q_{k'k}(U'|\theta')}{\pi(k,\theta)p_{kk'}q_{kk'}(U|\theta)}\left|\frac{\partial g_{kk'}(\theta,U)}{\partial \theta\partial U}\right|\right\}.
\end{aligned}
\end{equation}

\section{Model Specification}\label{sec:Model}
\subsection{The dynamical structure function}
  We consider a network of $p$ measurable nodes, whose number of hidden nodes is unknown. The network can be described by a DSF as follows~\cite{yuan,yuan1}:
\begin{equation}
\begin{aligned}
Y = Q(q;\theta)Y + P(q;\theta)U + H(q;\theta)E.\\
\end{aligned}
\label{DSF}
\end{equation}
where $q$ denotes the time shift operator ($y(t+1)=qy(t)$). $Y\in \mathbb{R}^p$ are measurable nodes. $U\in \mathbb{R}^m$ are inputs. $E\in \mathbb{R}^q$ are i.i.d. Gaussian noise. $\theta$ are model parameters. 

$Q$, $P$ and $H$ are transfer matrices, each element of which is a transfer function, indicating that the network is a causal system. Matrix $Q$ implies the connectivity among observable nodes. Its transfer functions are strictly proper and its diagonal elements are zero. $P$ and $H$ matrices relate inputs and process noise to nodes, respectively. The transfer functions of matrix $P$ are strictly proper whilst those of matrix $H$ are proper. The topology of the network (i.e. model structure) is reflected by the zero structure of these three matrices. For example, if $[Q]_{ij}$ is zero, the $j$th node does not control the $i$th node. $\mathcal{M}_k$ denotes model structures and $M_k$ represents the corresponding number of links. In particular, $\mathcal{M}_1$ represents the fully-connected topology. The internal dynamics of the network are described by the transfer functions. The order of the transfer functions is unknown, which is related to the number of hidden states and their internal connectivity.

The input-output map of the network is deduced based on the DSF as follows:
\begin{equation}
\begin{aligned}
Y = G_uU + G_eE.
\end{aligned}
\end{equation}  
where
$G_u=(I-Q)^{-1}P$ and $
G_e=(I-Q)^{-1}H.$
Identifiability of the networks depends on whether the input-output map is associated to a unique DSF. To ensure the inference problem is well-posed, additional constraints are imposed to the structure of transfer matrices.
\begin{prop}[Identifiability of DSF networks~\cite{qpmodel}] Given a $p\times (m+q)$ transfer matrix $G=[G_u, G_e]$, the corresponding DSF is unique if and only if $p-1$ elements in each column of $[Q, P, H]'$ are known, which uniquely specifies the component of $(Q,P,H)$ in the null space of $[G', I]$.
\end{prop}

A sufficient condition for network identifiability is that matrix $H$ is diagonal so that $p-1$ elements in each column of $[Q, P, H]'$ are known to be zero. In what follows, we make following assumptions so that no prior knowledge of matrix $P$ is required to guarantee network identifiability.
\begin{assump}
	Noise matrix $H$ is diagonal, monic ($\lim_{q\rightarrow \infty} H=I$) and minimal phase.
\end{assump}

The target networks we consider are sparse and stable. Hence, we make a further assumption regarding network properties.
\begin{assump}
	Transfer matrices, $Q$ and $P$ are stable and sparse.
\end{assump}

\subsection{The likelihood distribution}
After simple manipulations, the DSF in~\eqref{DSF} can be reformulated as:
\begin{equation}
\begin{aligned}
Y &= F_y(q;\theta)Y +F_u(q;\theta)U + E.
\end{aligned}
\label{pred}
\end{equation}
where
\begin{equation}
\begin{aligned}
F_u(q;\theta) &= H^{-1}P.\\
F_y(q;\theta) &=  I-H^{-1}(I-Q).\\
\end{aligned}
\end{equation}
 According to the assumptions, transfer matrices, $F_u$ and $F_y$ are also stable. More importantly, since $H$ is diagonal, $F_u$ and $F_y$ have the same zero structure as $P$ and $Q$, respectively.

Identifying the transfer functions of model~\eqref{pred} is non-trivial. Since the number of hidden states is unknown, estimating the order of transfer functions requires an exhaustive search of all possibilities, which is computationally prohibitive for large-scale networks. Additionally, imposing stable transfer matrices is problematic. To simplify the identification problem, we express model~\eqref{pred} in a non-parametric way. By doing so, the selection of model complexity is avoided and, more importantly, system stability can be promoted effectively. The dynamical system for the $i$th target node, is formulated below:
\begin{equation}
\begin{aligned}
y_i(t) &= \sum_{j=1}^p\sum_{k=1}^{\infty}h_{ij}^y(k)y_j(t-k)\\
&+ \sum_{j=1}^m\sum_{k=1}^{\infty}h_{ij}^u(k)u_j(t-k)+e_i(t).
\end{aligned}
\label{nonpred}
\end{equation}
where $h_{ij}^y$ and $h_{ij}^u$ are the impulse responses of transfer functions $[F_y]_{ij}$ and $[F_u]_{ij}$, respectively. $e_i(t)$ is i.i.d. Gaussian noise. The objective is to estimate the impulse responses.

For the implementation purpose, the impulse responses are truncated after sample time $T$. $T$ is set sufficiently large in order to catch the major dynamics of the impulse responses (i.e. $|h(k)|\approx 0$ for $k\geq T$). Assume the availability of time-series data collected from discrete time indices $1$ to $N$ for each node and input. For the $i$th target node with $\mathcal{M}_1$ and a single experiment, we define the following matrices and vectors. For other possible model structures, $\mathcal{M}_k$, the corresponding terms are defined in the same way.
\begin{equation} 
\begin{aligned}
&Y = \left[\begin{array}{c}y_i(N)\\\vdots \\y_i(T+1)\end{array}\right], W = \left[\begin{array}{c}w_1\\ \hline \vdots \\ \hline w_{p+m}\end{array}\right].\\
&\Phi = \left[\begin{array}{cc}\Phi_y&\Phi_u\end{array}\right].\\
&\Phi_y =\left[\begin{array}{ccc} y_1(N-1:N-T)&\cdots&y_p(N-1:N-T)\\ \vdots &\ddots&\vdots \\y_1(T:1)&\cdots&y_p(T:1)\end{array}\right].\\
&\Phi_u =\left[\begin{array}{ccc} u_1(N-1:N-T)&\cdots&u_m(N-1:N-T)\\ \vdots &\ddots&\vdots \\u_1(T:1)&\cdots&u_m(T:1)\end{array}\right].\\
& \sigma = E\{e_i(t)^2\}.\\
\end{aligned}
\label{para}
\end{equation}
where $Y\in R^{N-T}$ are time-series of the $i$th node. $W\in R^{T(p+m)}$ contain $p+m$ groups of impulse responses, each of which corresponds to a transfer function of $F_y$ or $F_u$. $\Phi \in R^{(N-T)\times T(p+m)}$ include time series of all the nodes and inputs. $\sigma$ is the noise variance. Note that the dimension of these quantities varies with respect to the model structure. For example, if node $j$ does not control node $i$, $w_j$ and $\Phi(:,T(j-1):Tj)$ must be removed from the corresponding vector and matrix. As a result, for model structure $\mathcal{M}_k$, $Y\in R^{N-T}$, $W\in R^{TM_k}$ and $\Phi \in R^{(N-T)\times TM_k}$.

Based on Bayes' rules, the likelihood distribution of the $i$th target node with $\mathcal{M}_k$  is:
\begin{equation}
\begin{aligned}
p(Y\big|W,\sigma,D,\mathcal{M}_k)
&=(2\pi\sigma)^{-\frac{N-T}{2}}\exp\left\{-\frac{1}{2\sigma}\|Y-\Phi W\|_2^2\right\}.
\end{aligned}
\label{like}
\end{equation}
where $D$ denotes the measurements of other nodes and inputs. To simplify the notation, $D$ is suppressed in the following discussion.

\subsection{The prior distributions}
Full Bayesian treatment deploys prior distributions for each random quantity to build up a hierarchical structure. The prior distributions reflect prior knowledge and assumptions of the networks.

Under the Bayesian paradigm, impulse responses are assumed to be independent Gaussian processes~\cite{nonp}. To impose stable impulse responses, the covariance function (kernel function) must be chosen carefully. It has been shown that Tuned/Correlated kernel (TC), Diagonal/Correlated kernel (DC) and second order stable spline kernel (SS) are all capable of characterizing a reproducing Hilbert space (RKHS) for stable impulse responses. They have been frequently applied in the system identification community~\cite{nonp1,kern}.  Hence, these three kernel functions are all considered in our framework. The prior distribution for $W$ is:
\begin{equation}
\begin{aligned}
p(W|\lambda,\beta,\mathcal{M}_k) = \prod_{i=1}^{M_k}  \mathcal{N}(w_i|0,\lambda_iK_i).\\
\end{aligned}
\end{equation} 
where $K_i\in\mathcal{R}^{T\times T}$, $\lambda=[\lambda_1,...,\lambda_{M_k}]'$, $\beta=[\beta_1,...,\beta_{M_k}]'$. Note that for DC kernel, $\beta_i$ is a row vector consisting of two elements whilst it is a scalar for TC and SS kernels.
\begin{equation}
\begin{aligned}
\left[K_i\right]_{ts} &= k(t,s;\beta_i),~\lambda_i\geq0,\\
k_{TC}(t,s;\beta_i) &= \begin{array}{cc}\beta_i^{max(t,s)},&\beta_i\in(0,1)\end{array},\\
k_{DC}(t,s;\beta_i) &= \begin{array}{cc}\beta_{i1}^{\frac{(t+s)}{2}}\beta_{i2}^{|t-s|},& \beta_{i1}\in(0,1),\ \beta_{i2}\in(-1,1)\end{array},\\
k_{SS}(t,s;\beta_i) &= \begin{array}{cc}\frac{\beta^{t+s+max(t,s)}}{2}-\frac{\beta^{3max(t,s)}}{6},& \beta_{i}\in(0,1)\end{array}.
\end{aligned}
\end{equation} 
In here, $\beta$ are hyperparameters of the kernel functions, which control the exponentially decaying rate of impulse responses~\cite{nonp2}. $\lambda$ are scale variables of the kernel functions. They play the role of  Automatic Relevance Determination (ARD) parameters that control sparsity~\cite{nonp}. If $\lambda_i$ approaches zero, the corresponding impulse responses $w_i$ are forced to zero, meaning they can be removed from the model.

Since $\sigma$ is non-negative, an Inverse-Gamma distribution is assigned as its conjugate prior. Without specific preference on $\sigma$, parameters $a_0$ and $b_0$ of the distribution are set to $0.001$, resulting in a non-informative prior:
\begin{equation}
\begin{aligned}
p(\sigma;a_0,b_0)= IG(\sigma;a_0,b_0)
=\frac{b_0^{a_0}}{\Gamma(a_0)}\sigma^{-a_0-1}e^{-\frac{b_0}{\sigma}},
\end{aligned}
\end{equation} 
where $\Gamma(\cdot)$ is the gamma function.

Instead of introducing equal probability for model structures, the prior distribution for $\mathcal{M}_k$ depends on the number of links, $M_k$. The cardinality of the set of all possible model structures equates to $|\mathcal{M}|=\sum_{i=0}^{M_1-1}\mathcal{C}(M_1-1,i)$ where $\mathcal{C}$ denotes the combination operator (i.e. $C(n,m)=\frac{n(n-1)\cdots(n-m+1)}{m!}$). The prior of $\mathcal{M}_k$ is a minor variant of the truncated Poisson distribution:
  \begin{equation}
  \begin{aligned}
  p(\mathcal{M}_k|\alpha) &=
  &=\frac{\alpha^{M_k}(M_k!)^{-1}}{\sum_{i=1}^{|\mathcal{M}|}\alpha^{M_i}(M_i!)^{-1}}.
  \end{aligned}
  \label{pM}
  \end{equation} 
  where $\alpha$ is the rate parameter of the Poisson distribution. Distribution~\eqref{pM} favours sparse topologies since higher probability is assigned to model structures with lower number of links. In addition, different topologies that have the same number of links are equally distributed.
  
Finally, hyperpriors are assinged to hyperparameters to complete the hierarchy. For non-negative hyperparameter $\lambda_i$, an  Inverse-Gamma distribution is applied as the conjugate prior: $p(\lambda_i;a_1,b_1)= IG(\lambda_i;a_1,b_1)$.  To impose sparsity, we set $a_1=2$ and $b_1=1$ so that the distribution has infinite variance (to support a wide domain) but puts most of weights over small values.
 
For hyperparameter $\beta_i$, a uniform distribution is employed as the prior:
\begin{equation}
\begin{aligned}
TC/SS:\ &p(\beta_i) = 1,~\beta_i\in(0,1),\\
DC:\ &p(\beta_i) = \frac{1}{2},~ \beta_{i1}\in(0,1), ~\beta_{i2}\in(-1,1).
\end{aligned}
\end{equation}  

The conjugate Gamma distribution is assigned to hyperparameter $\alpha$:
\begin{equation}
\begin{aligned}
p(\alpha;a_2,b_2)= Gamma(\alpha;a_2,b_2)=\frac{b_2^{a_2}}{\Gamma(a_2)}\alpha^{a_2-1}e^{-b_2\alpha}.
\end{aligned}
\end{equation}  
where $\alpha$ is equal to the mean of the Poisson distribution. To promote sparse topologies, we set $a_2=0.1$ and $b_2=1$ so that $p(\alpha;a_2,b_2)$ approaches infinity at $\alpha=0$. Nevertheless, since parameters $a_2$ and $b_2$ are deep in the hierarchy, they have little impact on the model.

\subsection{The posterior distributions}
Consider that a heterogeneous dataset contains $L$ independent time series of the target network, which are collected under different experimental conditions.  It is reasonable to assume that the internal dynamics of the network may vary with experimental conditions but the network topology remains unchanged. Therefore, impulse responses under different experimental conditions are independently distributed.

  Based on Bayes' rules and after completing squares, the posterior distribution of DSF~\eqref{nonpred} is as follows:
   \begin{equation}
 \begin{aligned}
 &p(\mathcal{M}_k,W,\beta,\lambda,\sigma,\alpha|Y)\\
 &\varpropto p(Y|W,\sigma,\mathcal{M}_k)\left[\prod_{i=1}^Lp(W_i|\beta,\lambda,\sigma,\mathcal{M}_k)\right]p(\sigma)\\
 &\ \ \ \ p(\beta|\mathcal{M}_k)p(\lambda|\mathcal{M}_k)p(\mathcal{M}_k|\alpha)p(\alpha)\\
 &\varpropto\left[\prod_{j=1}^L(2\pi\sigma_j)^{-\frac{N_j-T}{2}}\exp\left\{-\frac{1}{2}Y_j'(\sigma_j I+\Phi_j \Lambda K\Phi_j')^{-1}Y_j\right\}\right.\\
 &\times|2\pi\Lambda K|^{-\frac{1}{2}}\exp\left\{-\frac{1}{2}(W_j-\mu_j)'\Sigma_j^{-1}(W_j-\mu_j)\right\}\\
 &\left.\times\sigma_j^{-a_0-1}\exp\left\{-\frac{b_0}{\sigma_j}\right\}\right]\alpha^{a_2-1}\exp\{-b_2\alpha\}\\
 &\times\frac{\alpha^{M_k}(M_k!)^{-1}}{\sum_{i=1}^{|\mathcal{M}|}\alpha^{M_i}(M_i!)^{-1}}\prod_{i=1}^{M_k}\frac{b_1^{a_1}}{2\Gamma(a_1)}\lambda_i^{-a_1-1}\exp\left\{-\frac{b_1}{\lambda_i}\right\}.
 \end{aligned}
 \label{fullB}
 \end{equation}  
 where subscript $j$ denotes the index of experiments. Note that hyperparameters are shared by different experiments, reflecting the initial belief that the variation of system dynamics is limited. The number of measurements of the $j$th experiment is $N_j$.
 \begin{equation}
 \begin{aligned}
 K &= blkdiag\{K_1,\cdots,K_{M_k}\},~\Lambda= diag\{\lambda\}\otimes I_T,\\
 \Sigma_j^{-1} &= \frac{1}{\sigma_j}\Phi_j'\Phi_j+(\Lambda K)^{-1},~
 \mu_j = \frac{1}{\sigma_j}\Sigma_j\Phi_j'Y_j.\\
 \end{aligned}
 \end{equation}  
According to the full Bayesian model~\eqref{fullB}, the conditional posterior distributions of the random variables are listed below for further discussion:
  \begin{equation}
 \begin{aligned}
 p(W|\beta,\lambda,\sigma,\alpha,\mathcal{M}_k,Y)&=\prod_{j=1}^L\mathcal{N}(W_j|\mu_j,\Sigma_j),\\
 p(\sigma|W,\beta,\lambda,\alpha,\mathcal{M}_k,Y)&=\prod_{j=1}^LIG(\sigma_j;a_{\sigma_j},b_{\sigma_j}),\\
 p(\alpha|W,\beta,\lambda,\sigma,\mathcal{M}_k,Y)&\varpropto \frac{\alpha^{a_2-1+M_k}(M_k!)^{-1}}{\sum_{i=1}^{|\mathcal{M}|}\alpha^{M_i}(M_i!)^{-1}}e^{-b_2\alpha},
 \end{aligned}
 \label{pd}
 \end{equation}  
 where
  \begin{equation}
 \begin{aligned}
 a_{\sigma_j}=a_0+\frac{N_j-T}{2},~
 b_{\sigma_j}=b_0+\frac{\|Y_j-\Phi_j W_j\|_2^2}{2}.
 \end{aligned}
 \end{equation}  
 By marginalizing $W$ out from the full Bayesian model~\eqref{fullB}, the reduced joint posterior distribution of $\mathcal{M}_k$, $\beta$ and $\lambda$ is as follows:
  \begin{equation}\small{
 \begin{aligned}
 &p(\beta,\lambda,\mathcal{M}_k|\sigma,\alpha,Y)\\
 &\varpropto \prod_{j=1}^L|\sigma_j I+\Phi_j\Lambda K\Phi_j'|^{-\frac{1}{2}}\exp\{-\frac{1}{2}Y_j'(\sigma_j I+\Phi_j\Lambda K\Phi_j')^{-1}Y_j\}\\
 &\times\frac{\alpha^{M_k}(M_k!)^{-1}}{\sum_{i=1}^{|\mathcal{M}|}\alpha^{M_i}(M_i!)^{-1}}\prod_{i=1}^{M_k}\frac{b_1^{a_1}}{2\Gamma(a_1)}\lambda_i^{-a_1-1}\exp\left\{-\frac{b_1}{\lambda_i}\right\}.
 \end{aligned}
 \label{marg}}
 \end{equation}

\section{Network Inference using RJMCMC}\label{sec:INF}
\subsection{Sampler with fixed topology}
With the full Bayesian model, we are interested in the posterior distribution of model structures, $p(\mathcal{M}_k|Y)$, by which we can determine the most likely network topology. Given the estimated model structure, we can evaluate impulse responses and noise variance by calculating their expectation, which requires exploring $p(W|\mathcal{M}_k,Y)$ and $p(\sigma|\mathcal{M}_k,Y)$. However, distributions $p(\mathcal{M}_k|Y)$, $p(W|\mathcal{M}_k,Y)$ and $p(\sigma|\mathcal{M}_k,Y)$ are intractable since they need to perform high-dimensional integrals of the nonlinear Bayesian model in~\eqref{fullB}. To solve the problem, numerical sampling methods are applied in our framework. 

To begin with, assume that the topology of the target network is known \textit{a priori} (e.g. $\mathcal{M}_k$). Since the dimension of random variables is unchanged, a traditional MCMC algorithm is sufficient to draw samples from the distribution~\eqref{fullB}. To improve the convergence property, some random quantities are marginalized out in certain sampling steps. The resulting MH-within-PCG sampler is designed following the rules of marginalization, permutation and trimming. The sampler is further modified to explore the network with unknown topology.

Gibbs sampling is accepted to construct the basic sampler (Sampler $1$) for drawing samples.

\begin{center}
	\begin{tabular}{l}
		\hline
		\textbf{Sampler 1}: Blocked Gibbs sampler\\ 
		\hline
		1: Sample $p(W^{t+1}|\beta^t,\lambda^t,\sigma^t,\alpha^t,\mathcal{M}_k,Y)$\\
		2: Sample $p(\beta^{t+1},\lambda^{t+1}|W^{t+1},\sigma^t,\alpha^t,\mathcal{M}_k,Y)$\\
		3: Sample $p(\sigma^{t+1}|W^{t+1},\beta^{t+1},\lambda^{t+1},\alpha^t,\mathcal{M}_k,Y)$\\
		4: Sample $p(\alpha^{t+1}|W^{t+1},\beta^{t+1},\lambda^{t+1},\sigma^{t+1},\mathcal{M}_k,Y)$ \\
		\hline
	\end{tabular}
\label{S1}
\end{center}

Since the distributions of steps 2 and 4 are known up to a normalization constant, these two sampling steps should be replaced by the MH method, leading to a MH-within-Gibbs sampler (Sampler 2). Such a replacement maintains the invariant distribution because no marginal distributions are called in the sampler.

\begin{center}
	\resizebox{\columnwidth}{!}{
		\begin{tabular}{l}
			\hline
			\textbf{Sampler 2}: MH-within-Gibbs sampler\\ 
			\hline
			1: Sample $p(W^{t+1}|\beta^t,\lambda^t,\sigma^t,\alpha^t,\mathcal{M}_k,Y)$\\
			2: Sample $p(\beta^{t+1},\lambda^{t+1}|W^{t+1},\sigma^t,\alpha^t,\mathcal{M}_k,Y)$ using MH \\
			3: Sample $p(\sigma^{t+1}|W^{t+1},\beta^{t+1},\lambda^{t+1},\alpha^t,\mathcal{M}_k,Y)$\\
			4: Sample $p(\alpha^{t+1}|W^{t+1},\beta^{t+1},\lambda^{t+1},\sigma^{t+1},\mathcal{M}_k,Y)$ using MH \\
			\hline
		\end{tabular}
	}
	\label{S2}
\end{center}

As indicated in~\eqref{marg}, one can marginalize $W$ out from distribution $p(\beta,\lambda|W,\sigma,\alpha,\mathcal{M}_k,Y)$ in step 2 of Sampler 2. According to the rule of marginalization, the reduced sampling step is followed immediately by a direct draw from the  conditional distribution of $w$, resulting in a MH-within-PCG sampler (Sampler 3). The quantity  that is not conditioned on in the next step is labelled with an asterisk.

\begin{center}
	\resizebox{\columnwidth}{!}{
		\begin{tabular}{l}
			\hline
			\textbf{Sampler 3}: Marginalization\\ 
			\hline
			1: Sample $p(W^{\star}|\beta^t,\lambda^t,\sigma^t,\alpha^t,\mathcal{M}_k,Y)$\\
			2: Sample $p(\beta^{t+1},\lambda^{t+1}|\sigma^t,\alpha^t,\mathcal{M}_k,Y)$ using MH \\
			3: Sample $p(W^{t+1}|\beta^{t+1},\lambda^{t+1},\sigma^{t},\alpha^t,\mathcal{M}_k,Y)$\\
			4: Sample $p(\sigma^{t+1}|W^{t+1},\beta^{t+1},\lambda^{t+1},\alpha^t,\mathcal{M}_k,Y)$ \\
			5: Sample $p(\alpha^{t+1}|W^{t+1},\beta^{t+1},\lambda^{t+1},\sigma^{t+1},\mathcal{M}_k,Y)$ using MH\\
			\hline
		\end{tabular}
	}
	\label{S3}
\end{center}

To further simply Sampler 3, permutation and trimming are applied. Note that in order to maintain the invariant distribution, steps 2 and 3 can neither be separated nor swapped. Since the sampled $W$ in step 1 is not conditioned on in step 2, step 1 is trimmed out safely. The resulting sampler is presented in Sampler 4.

\begin{center}
	\resizebox{\columnwidth}{!}{
		\begin{tabular}{l}
			\hline
			\textbf{Sampler 4}: Permutation and Trimming\\ 
			\hline
			1: Sample $p(\beta^{t+1},\lambda^{t+1}|\sigma^t,\alpha^t,\mathcal{M}_k,Y)$ using MH \\
			2: Sample $p(W^{t+1}|\beta^{t+1},\lambda^{t+1},\sigma^{t},\alpha^t,\mathcal{M}_k,Y)$ \\
			3: Sample $p(\sigma^{t+1}|W^{t+1},\beta^{t+1},\lambda^{t+1},\alpha^t,\mathcal{M}_k,Y)$\\
			4: Sample $p(\alpha^{t+1}|W^{t+1},\beta^{t+1},\lambda^{t+1},\sigma^{t+1},\mathcal{M}_k,Y)$ using MH\\
			\hline
		\end{tabular}
	}
	\label{S4}
\end{center}

The sampling steps of Sampler 4 can be rearranged in many other ways. For example, Sampler 4 can be modified as $4\rightarrow3 \rightarrow 1 \rightarrow 2$. However, not all the arrangements are valid. For instance, sequence $2 \rightarrow 1 \rightarrow 3 \rightarrow 4$  derived from trimming out step 3 of Sampler 3 is incorrect because it violates the rule of trimming. The point is that for a PCG sampler, its sampling steps cannot be replaced by their MH counterparts directly. Otherwise, the invariant distribution may be changed. To avoid this type of error, it is necessary to deduce a MH-within-PCG sampler step-by-step.

According to~\eqref{pd}, steps 2 and 3 of Sampler 4 can be implemented directly. Only steps 1 and 4 require further discussion. Since these two steps employ the MH approach, the proposal distributions for Markov states are first designed to produce candidate samples.

Since $\lambda$ are non-negative, a truncated Gaussian distribution is adopted to draw the proposal. For a Gaussian distribution of $\theta$ with mean $\mu_0$ and variance $\sigma_0$, its truncated probability density function on $(l,u)$ is:
 \begin{equation}
\begin{aligned}
&p_{\mathcal{N}}(\theta;\mu_0,\sigma_0,l,u)=\frac{f(\frac{\theta-\mu_0}{\sigma_0})}{\sigma_0\left[F(\frac{u-\mu_0}{\sigma_0})-F(\frac{l-\mu_0}{\sigma_0})\right]},
\end{aligned}
\end{equation}  
where
\begin{equation}
\begin{aligned}
f(x) &= \frac{1}{\sqrt{2\pi}}e^{-\frac{x^2}{2}},\\
F(x) &= \frac{1}{2}\left[1+erf(\frac{x}{\sqrt{2}})\right],\\
erf(x) &= \frac{2}{\sqrt{\pi}}\int_{0}^xe^{-t^2}dt.
\end{aligned}
\end{equation}  

The proposal distribution for $\lambda$ is $q(\lambda|\lambda^t)=\prod_{i=1}^{M_k}p_{\mathcal{N}}(\lambda_i;\lambda_i^t,0.05,0,+\infty)$. In order to avoid the high rejection rate, the proposed $\lambda^{p}$ only deviate from the current state $\lambda^t$ with small variance.

For hyperparameter $\beta$, the proposal of each element is drawn from the distribution independently as follows. For a random variable $\theta\in(l,u)$ with its expected value $\bar{\theta}$:
\begin{equation}
\begin{aligned}
p_U(\theta;\bar{\theta},l,u,\varepsilon)=\left\{\begin{array}{ll}U(\bar{\theta}-\frac{\varepsilon}{2},\bar{\theta}+\frac{\varepsilon}{2}) &l+\frac{\varepsilon}{2}<\bar{\theta}< u-\frac{\varepsilon}{2}\\ U(l,l+\varepsilon)& \bar{\theta}\leq l+\frac{\varepsilon}{2}\\U(u-\varepsilon,u)& \bar{\theta}\geq u-\frac{\varepsilon}{2}\end{array}\right. .
\end{aligned}
\end{equation}
where $U(a,b)$ is the uniform distribution on $(a,b)$. $\varepsilon$ is the selection window for sampling. Hence, the proposal distribution for $\beta_i$ is:
\begin{equation}
\begin{aligned}
&TC/SS:q(\beta_i|\beta_i^t)= p_U(\beta_i;\beta_i^t,0,1,0.1)\\
&DC:q(\beta_i|\beta_i^t)=p_U(\beta_{i1};\beta_{i1}^t,0,1,0.1)p_U(\beta_{i2};\beta_{i2}^t,-1,1,0.1).
\end{aligned}
\end{equation}

According to 'detailed balance', the acceptance probability for step 1 is calculated as follows:
\begin{equation}
\begin{aligned}
&A_U(\beta^{p},\lambda^{p}|\beta^t,\lambda^t)\\
&=\min\left\{1,\frac{p(\beta^{p},\lambda^{p}|\sigma^t,\alpha^t,\mathcal{M}_k,Y)q(\beta^t|\beta^{p})q(\lambda^t|\lambda^{p})}{p(\beta^t,\lambda^t|\sigma^t,\alpha^t,\mathcal{M}_k,Y)q(\beta^{p}|\beta^t)q(\lambda^{p}|\lambda^t)}\right\}\\
&=\min\left\{1,r_U(\beta^p,\lambda^p|\beta^t,\lambda^t)\right\}.\\
\end{aligned}
\end{equation}
where
\begin{equation}{\footnotesize
\begin{aligned}
&r_U(\beta^p,\lambda^p|\beta^t,\lambda^t)\\
 &=\left[\prod_{j=1}^L \frac{\exp\{-\frac{1}{2}Y_j'(\sigma_j^t I+\Phi_j\Lambda^p K^p\Phi_j')^{-1}Y\}|\sigma_j I+\Phi_j\Lambda^p K^p\Phi_j'|^{-\frac{1}{2}}}{\exp\{-\frac{1}{2}Y_j'(\sigma_j^t I+\Phi_j\Lambda^t K^t\Phi_j')^{-1}Y_j\}|\sigma_j I+\Phi_j\Lambda^t K^t\Phi_j'|^{-\frac{1}{2}}}\right]\\
&\times\prod_{i=1}^{M_k}\left(\frac{\lambda_i^p}{\lambda_i^t}\right)^{-a_1-1}\exp\left\{\frac{b_1(\lambda_i^p-\lambda_i^t)}{\lambda_i^p\lambda_i^t}\right\}\frac{1+erf(\frac{\lambda_i^t}{\sqrt{2}\sigma_0})}{1+erf(\frac{\lambda_i^p}{\sqrt{2}\sigma_0})}.
\end{aligned}}
\end{equation}  

In practice, the variance of the proposal distributions (i.e. $\sigma_0$ and $\varepsilon$) is tuned during inference so that the acceptance accounts for $40\%$ of the total iterations, according to a heuristic rule in~\cite{Mon}.

The MH sampling step for $\alpha$ in Sampler 4 is designed in the same way. The proposal is drawn from a Gamma distribution: $q(\alpha|\alpha^t)\varpropto\alpha^{a_2-1+M_k}e^{-(1+b_2)\alpha}$. Therefore, the acceptance probability is:
\begin{equation}
\begin{aligned}
A(\alpha^{p}|\alpha^t)&=\min\left\{1,\frac{e^{-\alpha^t}\sum_{i=1}^{|\mathcal{M}|}(\alpha^t)^{M_i}(M_i!)^{-1}}{e^{-\alpha^p}\sum_{i=1}^{|\mathcal{M}|}(\alpha^p)^{M_i}(M_i!)^{-1}}\right\}.
\end{aligned}
\end{equation}

Note that step 4 is independent on the other sampling steps. That is because hyperparameter $\alpha$ is only related to $\mathcal{M}_k$ that is pre-fixed in this section. As a result, step 4 can be removed from the sampler without affecting the convergence property. However, if $\mathcal{M}_k$ needs to be sampled (unknown topology), step 4 must be retained in the sampler.

\subsection{Sampler with unknown topology}

If  the network topology is unknown, $\mathcal{M}_k$ is treated as a random variable and needs to be sampled for topology detection. To sample from~\eqref{fullB}, a blocked Gibbs sampler (Sampler 5) is applied as the last section.
Since the dimension of  $W$, $\beta$ and $\lambda$ is dependent on $\mathcal{M}_k$, these random variables are grouped together.

\begin{center}
		\begin{tabular}{l}
			\hline
			\textbf{Sampler 5}: Blocked Gibbs sampler\\ 
			\hline
			1: Sample $p(W^{t+1},\beta^{t+1},\lambda^{t+1},\mathcal{M}_k^{t+1}|w^{t+1},\sigma^t,\alpha^t,Y)$ \\
			2: Sample $p(\sigma^{t+1}|W^{t+1},\beta^{t+1},\lambda^{t+1},\alpha^t,\mathcal{M}_k^{t+1},Y)$\\
			3: Sample $p(\alpha^{t+1}|W^{t+1},\beta^{t+1},\lambda^{t+1},\sigma^{t+1},\mathcal{M}_k^{t+1},Y)$ \\
			\hline
		\end{tabular}
	\label{S5}
\end{center}

Since the distributions of steps 1 and 3 in Sampler 5 cannot be sampled directly, these two steps are implemented using the MH method, leading to a MH-within-Gibbs sampler (Sampler 6).

\begin{center}
		\resizebox{\columnwidth}{!}{
	\begin{tabular}{l}
		\hline
		\textbf{Sampler 6}: MH-within-Gibbs sampler\\ 
		\hline
		1: Sample $p(W^{t+1},\beta^{t+1},\lambda^{t+1},\mathcal{M}_k^{t+1}|\sigma^t,\alpha^t,Y)$ using MH \\
		2: Sample $p(\sigma^{t+1}|W^{t+1},\beta^{t+1},\lambda^{t+1},\alpha^t,\mathcal{M}_k^{t+1},Y)$ \\
		3: Sample $p(\alpha^{t+1}|W^{t+1},\beta^{t+1},\lambda^{t+1},\sigma^{t+1},\mathcal{M}_k^{t+1},Y)$ using MH\\
		\hline
	\end{tabular}
}
	\label{S6}
\end{center}

According to the rule of marginalization, after marginalizing $w$ out from step 1, one must sample $w$ immediately from its conditional distribution in the next step. The resulting MH-within-PCG sampler is shown in Sampler 7. 

\begin{center}
	\resizebox{\columnwidth}{!}{
		\begin{tabular}{l}
			\hline
			\textbf{Sampler 7}: MH-within-PCG\\ 
			\hline
			1: Sample $p(\beta^{t+1},\lambda^{t+1},\mathcal{M}_k^{t+1}|\sigma^t,\alpha^t,Y)$ using MH\\
			2: Sample $p(W^{t+1}|\beta^{t+1},\lambda^{t+1},\sigma^{t},\alpha^t,\mathcal{M}_k^{t+1},Y)$	 \\
			3: Sample $p(\sigma^{t+1}|W^{t+1},\beta^{t+1},\lambda^{t+1},\alpha^t,\mathcal{M}_k^{t+1},Y)$	\\
			4: Sample $p(\alpha^{t+1}|W^{t+1},\beta^{t+1},\lambda^{t+1},\sigma^{t+1},\mathcal{M}_k^{t+1},Y)$ using MH\\
			\hline
		\end{tabular}
	}
	\label{S7}
\end{center}

Since $\mathcal{M}_k$ is fixed in steps 2, 3 and 4 of Sampler 7, the dimension of their random variables is unchanged. As a result, the sampling steps of Sampler 4 can be applied here directly. Sampler 7 explores different model structures (topologies) via step 1. Since the sampling space of step 1 is composed of multiple subspaces of differing dimensionality, the sampler must be capable of traversing these subspaces in order to explore the Bayesian model sufficiently. Towards this point, the RJMCMC scheme is applied in this section.

To realize effective jumps between the parameter subspaces, different types of moves  are proposed for the Markov chain. Successful moves should both allow the Markov chain to visit all possible subspaces and promote reasonable acceptance probability. Motivated by this idea, we come up with three types of moves as follows:

\textbf{\textit{Birth Move}:}
The number of links in the next state, $M_k^{t+1}$ is one more than that of the current state (i.e. $M_k^{t+1}=M_k^t+1$). Furthermore, the zero structure of $\mathcal{M}_k^{t+1}$ and $\mathcal{M}_k^{t}$ only differs at one entry. For example, the Boolean structure of $\mathcal{M}_k^{t}$ is $[\begin{array}{ccc}1& 0& 0\end{array}]$ and that of $\mathcal{M}_k^{t+1}$ is $[\begin{array}{ccc}1& 0& 1\end{array}]$.

\textbf{\textit{Death Move}:}
The number of links in the next state, $M_k^{t+1}$ is one less than that of the current state (i.e. $M_k^{t+1}=M_k^t-1$). Furthermore, the zero structure of $\mathcal{M}_k^{t+1}$ and $\mathcal{M}_k^{t}$ only differs at one entry. For example, the Boolean structure of $\mathcal{M}_k^{t}$ is $[\begin{array}{ccc}1& 0& 1\end{array}]$ and that of $\mathcal{M}_k^{t+1}$ is $[\begin{array}{ccc}1& 0& 0\end{array}]$.

\textbf{\textit{Update Move}:}
The topology of the network is unchanged in the next state (i.e. $\mathcal{M}_k^{t+1}=\mathcal{M}_k^t$) but the other random variables are updated.

The birth and death moves of RJMCMC encourage a global search of the parameter subspaces, leading to a thorough exploration of network topology. The death move is equivalent to setting an ARD parameter $\lambda_i$ to zero whilst the birth move reverses this process by retrieving non-zero $\lambda_i$. As a result, the effect of ARD is maximally activated. The update move inherently infers internal dynamics of the network to interpret the dataset.

 Kernel-based methods often apply ARD for topology detection, where the sparsity profile of ADR parameters determines network topology. Nevertheless, due to local optimal solutions and numerical errors, the estimated ARD parameters are often not strictly zero. One can try different initial points to somehow avoid local optima  or employ certain model selection strategies (e.g. backward selection) to enforce sparsity. Nevertheless, since these schemes either pick up local and global optima equally likely or implement algorithms repeatedly, they raise computational cost and can be highly inefficient. 

 The main advantage of RJMCMC is that it explores the parameter space in a highly effective way. The jump proposed by RJMCMC is not always accepted. Rather, the acceptance probability involves the trade-off between data-fitting and sparsity penalties as, for example, $r_U$ contains the ratio of cost functions of different model structures: these cost functions are minimized in the kernel-based methods. Compared with the KEB approaches that mainly search one parameter space of the highest dimensionality ($\mathcal{M}_1$), RJMCMC does not necessarily step into all parameter subspaces, meaning that the Markov chain (that consists of accepted Markov states) only contains not all but a number of model structures embedded in the lower-dimensional subspaces. In addition, some model structures are only visited with low frequency, implying they are unlikely to be the ground truth. As a result, RJMCMC is able to focus on exploring other parameter subspaces whose corresponding model structures are closer to the ground truth. Therefore, many local optima of impulse responses and hyperparameters are avoided.  Compared with KEB, RJMCMC greatly increases inference accuracy and improves computational efficiency. 

To realize birth and death moves, the following algorithms (Algorithm~\ref{B} and \ref{D}) are proposed.
\begin{algorithm}[!]
	\caption{Birth Move}
	\label{B}
	\begin{algorithmic}[1]
		\State With probability $P_B$, choose Birth move.
		\State Select a node to be added to the current topology randomly by the Uniform distribution: $q_B(i|\mathcal{M}_k^t)=\frac{1}{M_1-M_k^t}$.		
		\State Draw proposals $\beta^p_i$ and $\lambda^p_i$ from $q_B(\beta_i,\lambda_i)=q_B(\beta_i)q_B(\lambda_i)$ with $\beta^t$ and $\lambda^t$ unchanged, where
		\begin{equation}
		\begin{aligned}
		q_B(\lambda_i)&=IG(\lambda_i;a_1,b_1)\\
		TC/SS:\ q_B(\beta_i) &= U(\beta_i;0,1)\\
		DC:\ q_B(\beta_i) &= U(\beta_{i1};0,1)U(\beta_{i2};-1,1)\\
		\end{aligned}
		\end{equation}
		\State Accept with probability $A_B$. Combine $\beta^p_i$ and $\lambda^p_i$ with $\beta^t$ and $\lambda^t$ to generate $\beta^{t+1}$ and $\lambda^{t+1}$ if accepted.
	\end{algorithmic}
\end{algorithm}

\begin{algorithm}[!]
	\caption{Death Move}
	\label{D}
	\begin{algorithmic}[1]
		\State With probability $P_D$, choose Death move.
		\State Select a node to be removed from the current topology randomly by the Uniform distribution: $q_D(i|\mathcal{M}_k^t)=\frac{1}{M_k^t-1}$ where the auto-regression terms are always retained.		
		\State Remove $\beta_i^t$ and $\lambda_i^t$ from $\beta^t$ and $\lambda^t$ respectively with other elements unchanged.
		\State Accept with probability $A_D$.
	\end{algorithmic}
\end{algorithm}

The acceptance probability for birth and death moves is calculated based on 'detailed balance':
\begin{equation}
\begin{aligned}
&A_B(\beta^{p},\lambda^{p},\mathcal{M}_k^{p}|\beta^t,\lambda^t,\mathcal{M}_k^t)\\
 &= \min\{1,r_B(\beta^{p},\lambda^{p},\mathcal{M}_k^{p}|\beta^t,\lambda^t,\mathcal{M}_k^t)\}\\
&A_D(\beta^{p},\lambda^{p},\mathcal{M}_k^{p}|\beta^t,\lambda^t,\mathcal{M}_k^t)\\
 &= \min\{1,r_D(\beta^{p},\lambda^{p},\mathcal{M}_k^{p}|\beta^t,\lambda^t,\mathcal{M}_k^t)\},
\end{aligned}
\end{equation}
where
\begin{equation}{\footnotesize
\begin{aligned}
&r_B(\beta^{p},\lambda^{p},\mathcal{M}_k^{p}|\beta^t,\lambda^t,\mathcal{M}_k^t)\\
 &=\left[\prod_{j=1}^{L}\frac{\exp\{-\frac{1}{2}Y_j'(\sigma_j^t I+\Phi_j \Lambda^pK^{p}\Phi_j')^{-1}Y_j\}}{\exp\{-\frac{1}{2}Y_j'(\sigma_j^t I+\Phi_j \Lambda^tK^{t}\Phi_j')^{-1}Y_j\}}\frac{|\sigma_j^t I+\Phi_j \Lambda^pK^{p}\Phi_j'|^{-\frac{1}{2}}}{|\sigma_j^t I+\Phi_j \Lambda^tK^{t}\Phi_j'|^{-\frac{1}{2}}}\right]\\
  &\times\frac{P_D}{P_B}\frac{\alpha^t(M_1-M_k^t)}{M_k^p(M_k^p-1)}\\
 &r_D(\beta^{p},\lambda^{p},\mathcal{M}_k^{p}|\beta^t,\lambda^t,\mathcal{M}_k^t)=r_B^{-1}(\beta^t,\lambda^t,\mathcal{M}_k^t|\beta^{p},\lambda^{p},\mathcal{M}_k^{p}).
\end{aligned}}
\end{equation}

Finally, the update move is shown in Algorithm~\ref{U}. Since the topology is fixed, the proposal distributions and acceptance probability are exactly the same with those of Sampler 4.

\begin{algorithm}[!]
	\caption{Update Move}
	\label{U}
	\begin{algorithmic}[1]
		\State With probability $P_U$, choose Update move.
		\State Propose $\beta^p$ and $\lambda^p$ using the Uniform distribution and truncated Gaussian distributions, respectively.	
		\State Accept with probability $A_U$.
	\end{algorithmic}
\end{algorithm}

Note that the probability of three moves depends on $\mathcal{M}_k^t$ as follows:
\begin{equation}
\begin{aligned}
P_B &= \left\{\begin{array}{cc}0.3&1<M_k^t<M_1\\0&M_k^t=M_1\\0.6&M_k^t=1\end{array}\right.,\\
P_D &= \left\{\begin{array}{cc}0.3&1<M_k^t<M_1\\0.6&M_k^t=M_1\\0&M_k^t=1\end{array}\right.,\\
P_U &= 1-P_B-P_D.
\end{aligned}
\end{equation}

To conclude, Algorithm~\ref{final} presents network inference using RJMCMC.

\begin{algorithm}[!]
	\caption{RJMCMC for network inference}
	\label{final}
	\begin{algorithmic}[1]
		\State Initialize $W^0$, $\beta^0$, $\lambda^0$, $\sigma^0$, $\alpha^0$, $\mathcal{M}_k^0$.
		\For  {$t=1:t_{max}$}
		\State Sample $P_{move}$ from $U(0,1)$.
		\If {$P_{move}\leq P_B$}
		\State Execute Birth Move (Algorithm~\eqref{B}).
		\ElsIf {$P_{move}\leq P_B+P_D$}
		\State Execute Death Move (Algorithm~\eqref{D}).
        \Else
        \State Execute Update Move (Algorithm~\eqref{U}).
		\EndIf
		\State Sample $W^t$ according to~\eqref{pd}.
		\State Sample $\sigma^t$ according to~\eqref{pd}.
		\State Sample $\alpha^t$ from $p(\alpha|W^t,\beta^t,\lambda^t,\sigma^t,\mathcal{M}_k^t)$ using step 4 of Sampler 4.
		\EndFor
		\State Store $\{W^t\}$, $\{\mathcal{M}_k^t\}$ and $\{\sigma^t\}$.
	\end{algorithmic}
\end{algorithm}

\subsection{Detection of topology and estimation of model parameters}

  Detection of network topology is based on the posterior distribution of model structures, $p(\mathcal{M}_k|Y)$. By the merit of RJMCMC, one can estimate the true distribution using the empirical distribution constructed by the samples:
  \begin{equation}
  \begin{aligned}
  P(\mathcal{M}_k=\mathcal{M}_i|Y) &=  \frac{1}{t_{max}}\sum_{t=1}^{t_{max}} \mathbf{1}_{\mathcal{M}_i}(\mathcal{M}_k^t),
  \end{aligned}
   \label{emp}
  \end{equation}
  where
   \begin{equation}
  \begin{aligned}
  \mathbf{1}_{x}(y) &= \left\{\begin{array}{cc}1&y=x\\0&y\neq x\end{array}\right..
  \end{aligned}
  \end{equation}
  
  Given the empirical distribution~\eqref{emp}, the most likely network topology is estimated based on maximum a posteriori (MAP): $\mathcal{M}_{opt}=\max_{\mathcal{M}_k}P(\mathcal{M}_k|Y)$.
  
  In biology, biologists often prefer to evaluate the probability of each possible link. The generated topology is fully connected and the confidence of links is measured by their probability. To achieve this, one can evaluate the probability of the link from node $j$ to node $i$ ($j\rightarrow i$) as follows:
  \begin{equation}
  \begin{aligned}
  P(j\rightarrow i|Y) &= \sum_{k=1}^{|\mathcal{M}|}P(j\rightarrow i,\mathcal{M}_k|Y) \\
  &=\frac{1}{t_{max}}\sum_{\mathcal{M}_q|j\rightarrow i\in\mathcal{M}_q}\sum_{t=1}^{t_{max}}\mathbf{1}_{\mathcal{M}_q}(\mathcal{M}_k^t),
  \end{aligned}
  \end{equation}
  where $j\rightarrow i\in\mathcal{M}_q$ means that link $j\rightarrow i$ is contained in topology $\mathcal{M}_q$.
  
  Finally, impulse responses and noise variance are estimated for model simulation and prediction as follows:
  \begin{equation}
  \begin{aligned}
  \hat{w} &=E(w|\mathcal{M}_k=\mathcal{M}_{opt},Y)= \frac{\sum_{t=1}^{t_{max}}\mathbf{1}_{\mathcal{M}_{opt}}(\mathcal{M}_k^t)w^t}{\sum_{t=1}^{t_{max}}\mathbf{1}_{\mathcal{M}_{opt}}(\mathcal{M}_k^t)}, \\
  \hat{\sigma} &=E(\sigma|\mathcal{M}_k=\mathcal{M}_{opt},Y)= \frac{\sum_{t=1}^{t_{max}}\mathbf{1}_{\mathcal{M}_{opt}}(\mathcal{M}_k^t)\sigma^t}{\sum_{t=1}^{t_{max}}\mathbf{1}_{\mathcal{M}_{opt}}(\mathcal{M}_k^t)}.
  \end{aligned}
  \end{equation}

\section{Simulation}\label{sec:Simulation}

 To compare our method with KEB inference approaches, we conducted two series of Monte Carlo simulations. Different kernel functions, including DC, SS and TC kernels were used for inference. KEB solves the following optimization problem.
More details can be found in~\cite{nonp}.
\begin{equation}
\begin{aligned}
\arg\min_{\sigma,\gamma,\beta} Y'(\sigma I+\Phi\Lambda K\Phi')^{-1}Y+\ln |\sigma I+\Phi\Lambda K\Phi'|,
\end{aligned}
\label{EB}
\end{equation}
where $\Lambda\in\mathbb{R}^{(N-T)\times T(p+m)}$ contains all ARD parameters whose sparsity profile determines network topology. To further improve the detection of zero ARD elements, the backward selection method is used~\cite{nonp}.
%	\begin{algorithm}[!]
%		\caption{KEB}
%		\label{alg3}
%		\begin{algorithmic}[1]
%			\State Solve \eqref{EB} with $\mathcal{M}_0$.
%			\State Set the threshold $R=asc\{\|w_1\|,...,\|w_{n+p}\|\}$.
%			\For {k=1:n+p-1}
%			\State $I = \{i|\|w_i\|\leq R_k\}$
%			\State Set $\gamma_i=0$ if $i\in I$ and calculate cost function, $f_k$.
%			\EndFor
%			\State $\mathcal{M}_{opt}=argmin_{\mathcal{M}_k}f_k$
%		\end{algorithmic}
%	\end{algorithm}

In the first part of simulations, random DSF networks were generated with different types of topologies (including a ring structure). They were simulated under various noise levels and inferred using time series data of various lengths. To investigate the algorithm performance when inferring real-world networks, our method was further tested on a synthetic gene regulatory network of the circadian clock of \textit{Arabidopsis thaliana}. The method was compared with iCheMA, a state-of-the-art approach that was developed to infer biological networks~\cite{Infm2}.

For DSF networks, two criteria are used to evaluate the performance of algorithms, namely True Positive Rate (TPR) and Precision (PREC). TPR shows the percentage of the true links in the ground truths that are successfully inferred. Precision (PREC) equates to the rate of the correct links over all the inferred links. TPR and PREC together indicate the accuracy of inferred networks. If TPR is low, the inferred network misses many true links, thus lacking of useful information; where PREC is low, the generated network is not reliable. To investigate the accuracy of estimated system dynamics, the identified models were applied to predict the validation dataset that was not used for inference. The prediction accuracy is measured based on the metric as follows.
\begin{equation}
fitness = 100\left(1-\frac{\|y-\hat{y}\|}{y-\bar{y}}\right),
\end{equation}
where $y$ are the validation data of a certain node, $\hat{y}$ are the predicted output and $\bar{y}$ are the mean of the validation data. The average of the fitness of all nodes is calculated for discussion.

For the gene regulatory network, the area under the receiver operating characteristic curve (AUROC) and the area under
the precision recall curve (AUPREC) are applied instead. These two criteria are widely used in systems biology to evaluate the accuracy of inferred biological networks. The receiver operating characteristic curve and the precision recall curve are plotted based on the link confidence. The areas under these curves are calculated. AUROC and AUPREC reveals similar information with TPR and PREC, respectively.

\subsection{Random DSF networks}

\begin{figure}
	\centering
	\includegraphics[width=0.9\columnwidth]{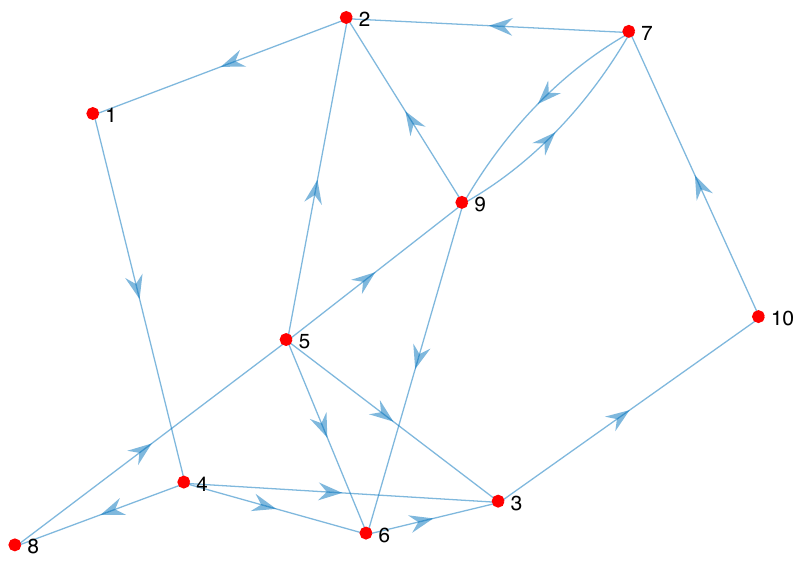}
	\caption{The structure of a randomly generated network. Solid lines with arrows represent links. Red circles denote nodes.}
	\label{random}
\end{figure}

$100$ networks were generated with random topologies and internal dynamics. All networks contained $15$ nodes, among which $10$ nodes were measured. Each node was independently driven by an input that was measured and process noise. To generate a state space model, a sparse stable matrix $A\in\mathcal{R}^{15\times15}$ was first yielded randomly using the function $sprandn(n,n,density)$ in Matlab. Matrix $A$ was guaranteed to be Hurwitz (i.e. no eigenvalue was outside the unit circle of the complex plane) using the brute-force strategy. No isolated nodes existed in the network. Figure~\ref{random} displays one example of the resulting networks.

To simulate the models, inputs and process noise were both i.i.d. white Gaussian signals. The variance of inputs was fixed to $1$ whilst that of process noise varied. The Signal-to-Noise ratio is defined as $SNR=10\log\frac{\sigma_u}{\sigma_e}$ where $\sigma_u$ and $\sigma_e$ are signal variance of inputs and noise, respectively. Only the first $10$ states of the models were measured. The truncation length of impulse responses was set to 20. The time series data were collected for inference  with various lengths between $45$ to $1000$.

The average TPR and PREC over $100$ trials are recorded in Table~\ref{M1}--\ref{M3}. In the best-case scenario (no procss noise), RJMCMC outperforms KEB methods in all cases. RJMCMC endowed with difference kernel functions present similar results. In particular, given sufficiently long time series ($\geqslant 65$), RJMCMC offers nearly perfect inference. In contrast, Kernel\_TC presents the weakest result. TPR of Kernel\_DC stays below $90\%$. Kernel\_SS is always outperformed by RJMCMC unless $85$ data points are used.

As $SNR$ decreases to $10dB$, RJMCMC exhibits a different performance using distinct kernel functions. RJMCMC\_DC and RJMCMC\_TC are both superior to RJMCMC\_SS and they show closely similar performance. In particular, RJMCMC\_DC and RJMCMC\_TC are always capable of producing reliable networks ($PREC\geqslant 97\%$). As the number of data points increases, they successfully capture most true links ($TPR\approx 90\%$). As with the previous simulation, in all cases KEB methods are no better than RJMCMC.

Under the worst-case scenario (no inputs), the performance improvement of RJMCMC is clearly evident compared with KEB methods. It is remarkable that PREC of RJMCMC\_DC and RJMCMC\_TC is always above $95\%$, indicating the inferred networks are highly reliable.
\begin{figure}
	\centering
	\includegraphics[width=1\columnwidth]{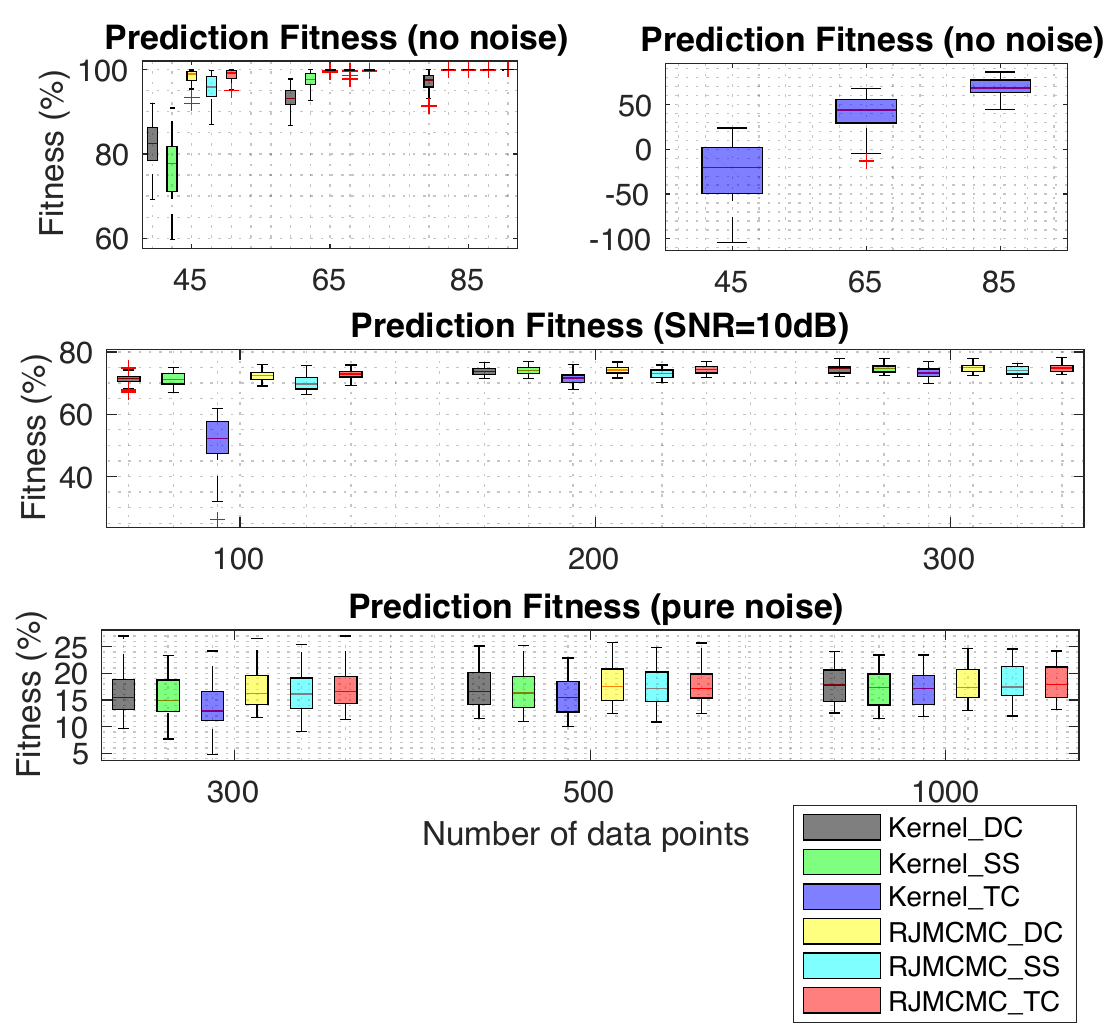}
	\caption{Prediction of randomly generated networks.}
	\label{fit_QP}
\end{figure}

The validation result is shown by the box plot in Figure~\ref{fit_QP}. The advantage of RJMCMC over KEB methods is evident under the best-case scenario whilst in other cases that is not so. The prediction accuracy of RJMCMC is slightly better than KEB methods for $SNR=10dB$ and pure noise cases. Nevertheless, Kernel\_TC always presents the weakest result.
\begin{table}[h!]
	\caption{Inference of random networks with no noise}
	\centering
	\resizebox{\columnwidth}{!}{
		\begin{tabular}{ |c|c|c|c|c|c|c|}
			\hline
			\multicolumn{7}{|c|}{No Noise}\\
			\hline
			\multirow{2}{2em}   &\multicolumn{2}{|c|}{45} &\multicolumn{2}{|c|}{65}&\multicolumn{2}{|c|}{85} \\
			\cline{2-7}
			&PREC &TPR&PREC&TPR&PREC&TPR\\
			\hline
			Kernel\_DC      &  91.2	&54.7&95.7	&73.5&	99.4	&84.2\\
			Kernel\_SS      &  82.8	&60.7&	89.6&	93.2&	99.9	&99.9\\
			Kernel\_TC      &50.1	&17.0&	76.9&	29.4&	91.2&	40.5\\
%			VI                   &100	&75.0&	99.7&	98.5&	100	&100\\
			RJMCMC\_DC        &99.4	&90.5&	100&	99.3&	100	&100\\
			RJMCMC\_SS        &93.4	&91.2&	100&	99.3&	100	&100\\
			RJMCMC\_TC        &99.6	&91.6&	100&	99.3&	100	&100\\
			\hline
		\end{tabular}
	}
	\label{M1}
\end{table}

% \begin{table}[h!]
% 	\caption{Inference of random networks with $20dB$ SNR}
% 	\centering
% 	\resizebox{\columnwidth}{!}{
% 	\begin{tabular}{ |c|c|c|c|c|c|c|  }
% 		\hline
% 		\multicolumn{7}{|c|}{20dB}\\
% 		\hline
% 		\multirow{2}{2em}   &\multicolumn{2}{|c|}{65} &\multicolumn{2}{|c|}{100}&\multicolumn{2}{|c|}{200} \\
% 		\cline{2-7}
% 		&PREC &TPR&PREC&TPR&PREC&TPR\\
% 		\hline
% 		Kernel\_SS      & 70.7&	86.7&	92.2&	91.6&	97.4&	96.5\\
% 		Kernel\_TC      &74.7&	28.7&	94.4&	46.6&	100&	69.6\\
% 		VI                   &100	&81.2&	99.4&	88.6&	100&	96.2\\
% 		\hline
% 	\end{tabular}
%}
% \end{table}

\begin{table}[h!]
	\caption{Inference of random networks with noise that has $10dB$ SNR}
	\centering 
	\resizebox{\columnwidth}{!}{
		\begin{tabular}{ |c|c|c|c|c|c|c|  }
			\hline
			\multicolumn{7}{|c|}{10dB}\\
			\hline
			\multirow{2}{2em}   &\multicolumn{2}{|c|}{100} &\multicolumn{2}{|c|}{200}&\multicolumn{2}{|c|}{300} \\
			\cline{2-7}
			&PREC &TPR&PREC&TPR&PREC&TPR\\
			\hline
			Kernel\_DC      & 91.9	&75.5&	97.2&	84.0&	98.0&	87.1\\
			Kernel\_SS      & 74.7	&82.9&	80.7&	88.8&	87.8&	89.0\\
			Kernel\_TC      &87.3	&47.5&	99.6	&67.6	&100&	74.2\\
			%VI                   &100&	68.2&	99.7	&81.3&	100	&86.4\\
			RJMCMC\_DC         &97.0&80.0&	98.2	&87.4&	98.5	&89.4\\
			RJMCMC\_SS         &68.3&85.8&	80.7	&90.1&	82.4	&93.2\\
			RJMCMC\_TC         &98.1&81.6&	98.0&88.5&	98.7	&90.1\\
			\hline
		\end{tabular}
	}
	\label{M2}
\end{table}

% \begin{table}[h!]
%    \caption{Inference of random networks with $0dB$ SNR}
% 	\centering
% 	\resizebox{\columnwidth}{!}{
% 	\begin{tabular}{ |c|c|c|c|c|c|c|  }
% 		\hline
% 		\multicolumn{7}{|c|}{0dB}\\
% 		\hline
% 		\multirow{2}{2em}   &\multicolumn{2}{|c|}{400} &\multicolumn{2}{|c|}{500}&\multicolumn{2}{|c|}{600} \\
% 		\cline{2-7}
% 		&PREC &TPR&PREC&TPR&PREC&TPR\\
% 		\hline
% 		Kernel\_SS      &78.9	&79.1&	80.6&	78.7&	81.7&	82.7\\
% 		Kernel\_TC      &90.2	&72.7	&94.8&75.3&92.9&	76.3		\\
% 		VI                   &100&	72.6&	100	&76.7&	100	&78.2\\
% 		\hline
% 	\end{tabular}
%}
% \end{table}

\begin{table}[h!]
	\caption{Inference of random networks with pure noise}
	\centering
	\resizebox{\columnwidth}{!}{
		\begin{tabular}{ |c|c|c|c|c|c|c|  }
			\hline
			\multicolumn{7}{|c|}{No Input}\\
			\hline
			\multirow{2}{2em}   &\multicolumn{2}{|c|}{300} &\multicolumn{2}{|c|}{500}&\multicolumn{2}{|c|}{1000} \\
			\cline{2-7}
			&PREC &TPR&PREC&TPR&PREC&TPR\\
			\hline
			Kernel\_DC      &81.0&71.5	&	85.6&	73.9&	97.2&	75.3\\
			Kernel\_SS      &66.4&68.8	&	80.5&	70.9&	82.0&	72.8\\
			Kernel\_TC      &81.6	&59.5	&89.2&66.2&93.4&71.3			\\
			%VI                   &100&56.5&	100	&65.0&	100	&74.9\\
			RJMCMC\_DC         &96.2&69.2&	98.8	&76.3&	97.5	&83\\
			RJMCMC\_SS         &78.6&74.3&	87.1	&76.5&	88.4	&84.5\\
			RJMCMC\_TC         &95.7&70.9&	96.7&76.7&	98.9	&81.8\\
			\hline
		\end{tabular}
	}
	\label{M3}
\end{table}

\subsection{Ring networks}

\begin{figure}
	\centering
	\includegraphics[width=0.7\columnwidth]{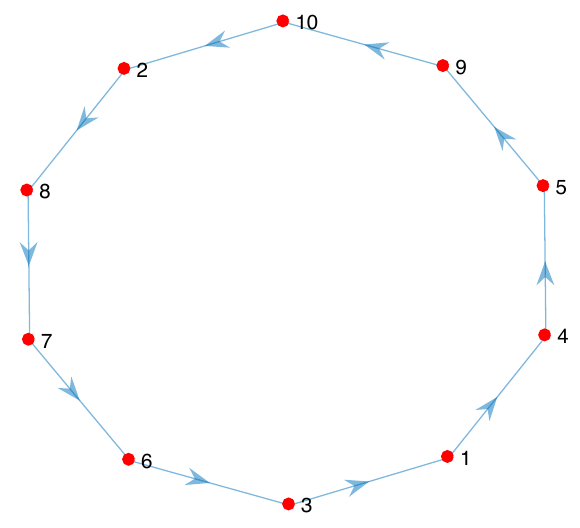}
	\caption{A network with the ring structure. Symbol '$\sim$' denotes the input signals.}
	\label{ring} 
\end{figure}

$100$ networks with the fixed ring structure (Figure~\ref{ring}) were generated and simulated following the same protocol of random DSFs. Each node was driven by independent process noise. Only one input entered the network through a single node. Since the network forms a closed feedback loop and is extremely sparse, it is more challenging to infer.

Table~\ref{R2} presents the inference result. Simulations indicate that RJMCMC is superior to KEB methods. In particular, RJMCMC\_DC and RJMCMC\_TC present the best results. PREC of these two cases always exceeds $90\%$ and increases to $98\%$ given $400$ data points, while only $1$ true link is missed. For KEB methods, either PREC or TPR is lower than the corresponding RJMCMC cases.

\begin{figure}
	\centering
	\includegraphics[width=1\columnwidth]{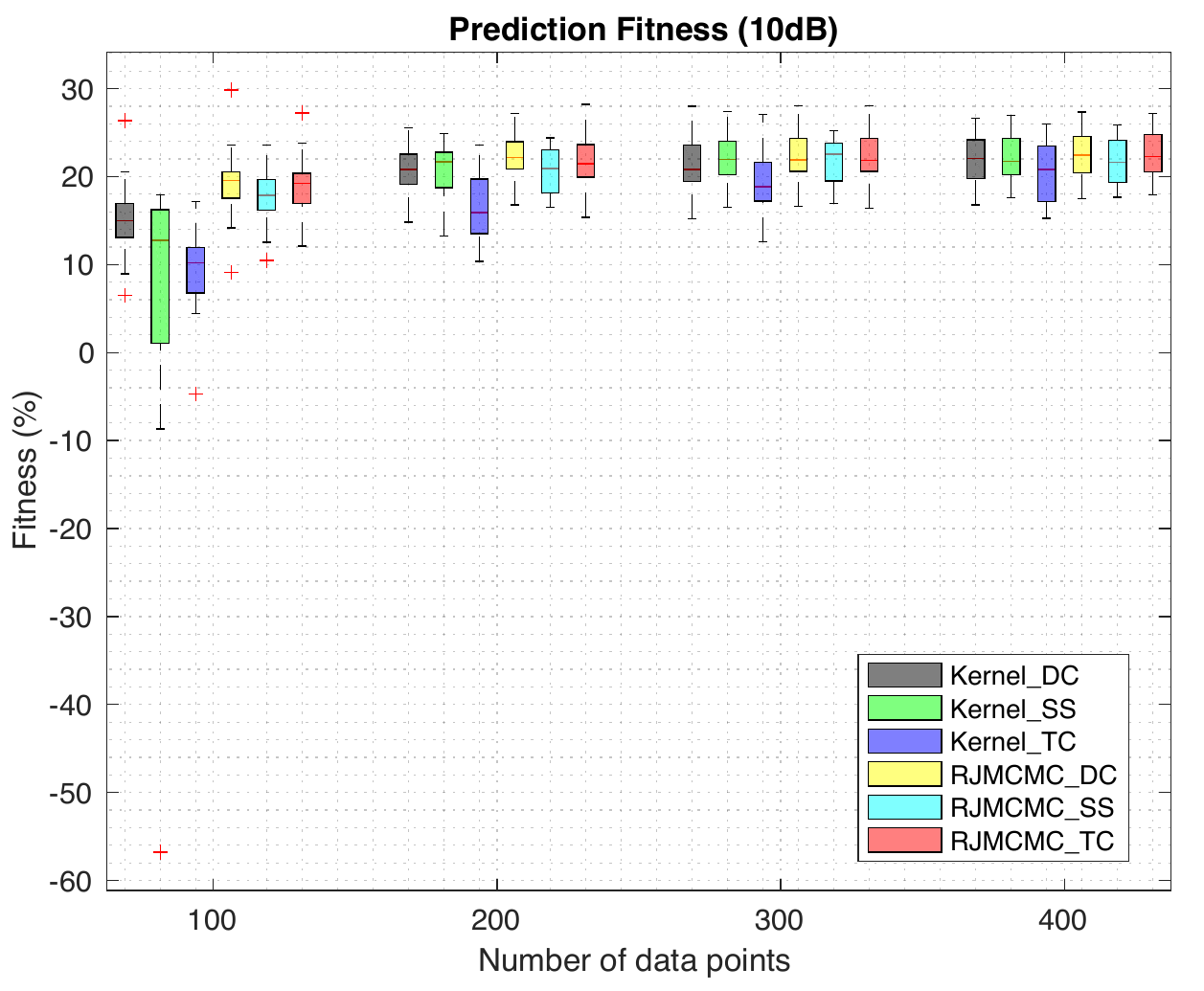}
	\caption{Prediction of ring networks.}
	\label{fit_Ring}
\end{figure}

The validation result in Figure~\ref{fit_Ring} shows that RJMCMC outperforms KEB methods, especially given lower number of data points.

%	\begin{table}[h!]
%		\caption{Inference of ring networks with pure noise}
%		\centering
%		\resizebox{\columnwidth}{!}{
%			\begin{tabular}{ |c|c|c|c|c|c|c|  }
%				\hline
%				\multicolumn{7}{|c|}{No Input }\\
%				\hline
%				\multirow{2}{2em}   &\multicolumn{2}{|c|}{300} &\multicolumn{2}{|c|}{500}&\multicolumn{2}{|c|}{1000} \\
%				\cline{2-7}
%				&PREC &TPR&PREC&TPR&PREC&TPR\\
%				\hline
%				Kernel\_SS      &61.4&82.5		&72.2&	85.0	&87.2	&85.0\\
%				Kernel\_TC      &73.8	&78	&88.3&80&95.5&	84		\\
%				VI                   &100&	73&	100	&82	&100&	87.5\\
%				\hline
%			\end{tabular}
%		}
%		\label{R1}
%	\end{table}

\begin{table}[h!]
	\caption{Inference of ring networks with $10dB$ SNR}
	\centering
	\resizebox{\columnwidth}{!}{
		\begin{tabular}{ |c|c|c|c|c|c|c|c|c|  }
			\hline
			\multicolumn{9}{|c|}{10dB }\\
			\hline
			\multirow{2}{2em}   &\multicolumn{2}{|c|}{100} &\multicolumn{2}{|c|}{200}&\multicolumn{2}{|c|}{300} &\multicolumn{2}{|c|}{400}\\
			\cline{2-9}
			&PREC &TPR&PREC&TPR&PREC&TPR&PREC&TPR\\
			\hline
			Kernel\_DC      &54.4&77.0		&75.8&	82.3	&85.0	&85.5&90.5&86.5\\
			Kernel\_SS      &41.2&78.5	&62.8&	83.5	&71.8	&90.0&71.3&89.0\\
			Kernel\_TC      &76.9	&19.3	&98.4&39.3&96.8&	56.0&97.6&67.8		\\
			%VI                   &100&	27.5&	100	&66.0	&100&	73.5&100&80.0\\
			RJMCMC\_DC         &92.0&67.3&	94.5	&79.3&	94.9	&85.5&98.1&88.8\\
			RJMCMC\_SS         &60.7&70.0&	81.1	&80.3&	81.2	&85.5&85.9&88.3\\
			RJMCMC\_TC         &91.2&66.3&	95.3	&80.8&	96.6	&86.8&98.0&89.3\\
			\hline
		\end{tabular}
	}
	\label{R2}
\end{table}

To conclude, RJMCMC tremendously improves the performance of KEB methods markedly in both inferring network topology and identifying internal dynamics. RJMCMC produces more accurate inference results. The generated networks are highly reliable and contain most true links in the ground truths.
% \begin{table}[h!]
% 		\caption{Inference of ring networks with $20dB$ SNR}
% 	\centering
% 	\resizebox{\columnwidth}{!}{
% 	\begin{tabular}{ |c|c|c|c|c|c|c|  }
% 		\hline
% 		\multicolumn{7}{|c|}{20dB}\\
% 		\hline
% 		\multirow{2}{2em}   &\multicolumn{2}{|c|}{65} &\multicolumn{2}{|c|}{100}&\multicolumn{2}{|c|}{200} \\
% 		\cline{2-7}
% 		&PREC &TPR&PREC&TPR&PREC&TPR\\
% 		\hline
% 		Kernel\_SS      &61.4&82.5		&72.2&	85.0	&87.2	&85.0\\
% 		Kernel\_TC      &73.8	&78	&88.3&80&95.5&	84		\\
% 		VI                   &100&	73&	100	&82	&100&	87.5\\
% 		\hline
% 	\end{tabular}
%}
% 	\label{R2}
% \end{table}

% \begin{table}
% 	\centering
% 	\resizebox{\columnwidth}{!}{
% 	\begin{tabular}{ |c|c|c|c|c|c|c|  }
% 		\hline
% 		\multicolumn{7}{|c|}{Unidentifiable DSF}\\
% 		\hline
% 		\multirow{2}{2em}   &\multicolumn{2}{|c|}{45} &\multicolumn{2}{|c|}{65}&\multicolumn{2}{|c|}{100} \\
% 		\cline{2-7}
% 		&PREC &TPR&PREC&TPR&PREC&TPR\\
% 		\hline
% 		Kernel\_SS      &51.9&	69.0	&58.5&68.6		&61.8	&66.5\\
% 		Kernel\_TC      &49.3	&34.3&42.2&43.6&50.2&39.3			\\
% 		VI\_TC            &50&19.4	&36.5	&16	&&	\\
% 		\hline
% 	\end{tabular}
%}
% \end{table}

\subsection{Synthetic circadian clock network}

In above two simulations, the ground truth models fall exactly in the proposed model class. Nevertheless, many of the real-world networks are nonlinear.  Under our framework, linear models are used as the approximation in order to deal with unmeasurable nodes. To check the effectiveness of our method, a synthetic model of the circadian clock (Millar 10~\cite{Millar10}), was employed for test. In addition, we compared our method with a state-of-the-art technique, iCheMA that has been shown to outperform many existing inference methods, including hierarchical Bayesian regression (HBR), LASSO and elastic net through Monte Carlo simulations on the Millar 10 model~\cite{Infm2}.

 Millar 10 describes a circadian clock consisting of $7$ genes along with their associated proteins, which amounts to $19$ nodes in total. The system is driven by light signals. The detailed mathematical model can be found in~\cite{Millar10}. The simulation aimed to produce synthetic microarray data. The time window for data collection was $44$ hours. The sampling frequency was $1$ hour: as a consequence, only $44$ data points were used for each trial. Most importantly, the protein data were not available for inference. Therefore, the network was inferred on the transcriptional level, describing the connectivity among $7$ clock genes. The model was simulated for four days of light-dark cycles (LD for one day) followed by three days of constant light (LL for one day). The simulation was repeated $50$ times. To avoid the transition due to the initial condition, the simulated data of the first two days were discarded. Time windows of LDLD (0h-44h), LDLL (24h-68h), LLLL (48h-92h) and steady state (72h-116h) were adopted for data collection. Considering only $44$ data points were available for inference, the length of truncated impulse responses was set to $10$. For the kernel methods, we resorted to~\cite{Infm1} to calculate the confidence of inferred links as $P(j\rightarrow i|Y)=\frac{\|w_j\|}{\|w\|}$.

The inference result is presented in Table~\ref{clock}. RJMCMC\_SS outperforms all the other methods in most cases. In particular, under time window LDLL, both AUPREC and AUROC of RJMCMC\_SS are above $70\%$. Since this time window contains richest light transitions, the inference result indicates that RJMCMC is able to infer complex system dynamics. More importantly, the inference accuracy of RJMCMC is markedly improved compared with KEB methods. KEB methods perform poorly in inferring Millar 10. The inferred networks are unreliable and most true links are missed.  iCheMA is only slightly superior to RJMCMC\_SS under time window LLLL and is outperformed by RJMCMC\_SS in other cases.

Simulations imply that our method is reliable when dealing with real-world networks, especially for the cases where full state measurements are unavailable. Therefore, our method can be applied under a wide range of contexts such as biological networks, power grids and communication systems.

\begin{table*}[t]
	\centering
	\caption{Inference results of the circadian clock model.}
	\begin{center}
		\resizebox{1.5\columnwidth}{!}{
			\begin{tabular}{|c|c|c|c|c|c|c|c|c|}
				\hline
				&\multicolumn{2}{|c|}{LDLD}&\multicolumn{2}{|c|}{LDLL}&\multicolumn{2}{|c|}{LLLL} &\multicolumn{2}{|c|}{Steady State}\\
				\hline
				&AUROC& AUPREC&AUROC& AUPREC &AUROC& AUPREC&AUROC&AUPREC \\
				\hline
				iCheMA &66.4 \% & 62.3\%& 65.4\%& 64.2\%&69.4\% &66.8\%&64.7\%&56.4\%\\
				Kernel\_DC    &54.9 \% & 45.4\%&63.1 \% & 55.3\% &53.6 \% & 39.4\% &48.9\%&35.0\%   \\
				Kernel\_SS    &51.3 \% & 38.8\%&63.3 \% & 51.6\% &54.2 \% & 39.7\% &51.7\%&37.4\%   \\
				Kernel\_TC    &48.6 \% & 36.5\%&55.5 \% & 43.7\% &51.7 \% & 38.9\% &47.4\%&35.5\%   \\
				RJMCMC\_DC    &64.8 \% & 61.1\%&68.7 \% & 66.7\% &61.8 \% & 57.2\% &58.2\%&54.0\%   \\
				RJMCMC\_SS    &69.4 \% & 63.8\%&76.5 \% & 73.4\% &66.3 \% & 62.7\% &64.6\%&62.8\%   \\
				RJMCMC\_TC    &68.0 \% & 61.8\%&72.2 \% & 68.9\% &61.2 \% & 57.0\% &59.1\%&54.5\%   \\
				\hline
			\end{tabular}
		}
	\end{center}
\label{clock}
\end{table*}

\section{Conclusion}\label{sec:Conclusion}
This paper combines kernel-based system identification methods and RJMCMC to infer sparse networks. DSF models are used to describe the target network so that the information of hidden nodes is encoded via transfer functions. The models are expressed in a non-parametric way. By doing so, inference can be conducted without prior knowledge of the number of hidden nodes and their connectivity. The kernel machine is used to impose stable impulse responses. To sufficiently explore the full Bayesian model, RJMCMC is applied to draw samples from the space that is composed of subspaces of different dimensionality. By traversing the subspaces, RJMCMC greatly improves the accuracy of topology detection. Monte Carlo simulations demonstrate our method superior to KEB methods. In particular, the proposed method achieves marked advantages over KEB when inferring synthetic biological networks.

Overall, the value of this approach is that it always generates reliable inference results and is robust to experimental conditions, including the number of data points, types of topologies and noise levels. Given a sufficient data source, our method is able to infer most true links and is applicable to a wide range of real-world networks, where full state measurements are not available. According to the simulations, our method can be used to study circadian clocks. For example, it can be applied to infer the $Ca^{2+}$ signalling network of \textit{Arabidopsis}.

The method does, however, have some limitations, as follows. The computational cost is heavy when dealing with large-scale networks. Furthermore, the method for identification of continuous time systems requires high sampling frequency and equal sampling steps. In addition, DSF is not well-defined for stochastic differential equations (SDE) since the Wiener process in the model is almost surely nowhere differentiable. Further work is required to extend the method to continuous time networks described by SDE.

% or


\appendices

%\section*{Acknowledgment}

\ifCLASSOPTIONcaptionsoff
  \newpage
\fi



%\bibliographystyle{IEEEtran}
%\begin{thebibliography}{1}
%\end{thebibliography}
\bibliographystyle{IEEEtran}
\bibliography{IEEEabrv,IEEEexample}
%\end{IEEEbiography}
%% insert where needed to balance the two columns on the last page with
%%\newpage
%% You can push biographies down or up by placing
%% a \vfill before or after them. The appropriate
%% use of \vfill depends on what kind of text is
%% on the last page and whether or not the columns
%% are being equalized.
%
%%\vfill
%
%% Can be used to pull up biographies so that the bottom of the last one
%% is flush with the other column.
%%\enlargethispage{-5in}
%
%

\end{document}